# Fourier-Transform Photothermal Spectroscopy with Transient-Lens Detection

ASHA S. THOMAS[1], TOLUWATOMI HAASTRUP[1], LUZHOU YU[1], MUYI ZHANG[2] AND JENNIFER P. OGILVIE[1*]

*[1]Department of Physics, University of Ottawa, 150 Louis-Pasteur Pvt, Ottawa, Ontario, K1N 6N5, Canada*
*[2]Department of Physics, University of Michigan, 450 Church St., Ann Arbor MI, USA 48109*
**jogilvi3@uottawa.ca*

**Abstract:** We demonstrate Fourier-transform photothermal spectroscopy using broadband supercontinuum excitation and transient-lens detection. The output of a Ti:sapphire oscillator is broadened in a photonic crystal fiber to generate the supercontinuum pump, which is converted into a phase-stable pulse pair with a controllable femtosecond delay using a birefringent interferometer. The absorption-induced heating from the pump generates a transient thermal lens in the sample, which is detected as a function of delay by an off-resonant 800 nm photothermal probe beam. Fourier transform of the interferometric photothermal response yields a frequency-domain signal corresponding to the sample absorptance spectrum weighted by the pump spectrum. After division by the measured pump spectrum, the reconstructed photothermal spectrum is compared with a direct frequency-domain linear absorption measurement to evaluate the accuracy of the reconstruction. This method combines Fourier-transform broadband excitation with thermal-lens readout, allowing broadband linear absorption spectra to be obtained from pump-induced refractive-index changes rather than from transmitted excitation intensity. We demonstrate the method on phloxine B.

## 1. Introduction

Photothermal spectroscopy comprises a class of highly sensitive techniques for characterizing the optical absorption and thermal properties of matter. These techniques detect the heat released through nonradiative relaxation following photoexcitation, which changes the local temperature and, in turn, other thermodynamic properties of the sample such as pressure and density [1]. Because the signal originates from absorbed energy, photothermal spectroscopy provides an indirect measure of absorption that is largely insensitive to scattering and reflection losses, with sensitivity reaching the single-molecule level [2–6].

The photothermal response can be detected in many ways, allowing the detection modality to be matched to the measurement and photothermal detection to be integrated with existing spectroscopic and microscopy techniques to extend their capabilities. The earliest photothermal absorption measurements used photoacoustic detection: in 1881, Bell moved a hearing tube along a dispersed spectrum, noting where each substance produced sound [7]. Optical detection was later achieved with the transient lens (TL) method, first observed as divergence of the beam from a gas laser with a liquid sample inside its cavity [8]. When a sample is excited by a Gaussian beam, the deposited heat follows the beam profile, with the temperature rise highest at the center and decreasing toward the edges. The resulting refractive-index gradient acts as a transient diverging lens, which can be detected through the defocusing of a non-resonant probe beam, whose strength scales with the absorbed optical energy [9]. Its high sensitivity to small temperature changes has made TL a widely used photothermal detection method [10]. Because nonradiative relaxation competes with radiative relaxation, TL is also complementary to fluorescence spectroscopy [9]: fluorescence probes the fraction of absorbed energy released radiatively, whereas TL detects the energy dissipated nonradiatively.

The sensitivity of TL detection has enabled absorption measurements at the single-molecule level [6] and underpins photothermal microscopy [4,5,11,12], which has been applied to single-molecule imaging [4,5]. Unlike fluorescence microscopy, photothermal imaging is label-free and therefore not limited by photobleaching or blinking. Photothermal microscopy has also been combined with super-resolution approaches to surpass the diffraction limit [12].

TL absorption spectroscopy is commonly performed at a single excitation wavelength, or by tuning the excitation wavelength to obtain spectral information [5,6,13]. Broadband photothermal spectroscopy has been demonstrated in the infrared using frequency-comb and Fourier-transform approaches. Dual-comb photothermal spectroscopy recovered near-infrared absorption spectra spanning ~10 nm through interferometric photothermal detection without scanning the excitation wavelength, reaching a detection limit of 8.7 ppm for acetylene in a hollow-core fiber with a sample volume of only 0.17 μL [14]. Razumtcev et al. combined a synchrotron infrared source spanning ~2.5–18.5 μm with optical- and fluorescence-detected photothermal IR microscopy, in which IR absorption is read out by a visible (532 nm) probe, giving spatial resolution beyond the IR diffraction limit. Absorption spectra were retrieved by Fourier transformation of photothermal interferograms [15,16] recorded with a step-scan interferometer. Each scan required up to 10 minutes, and the achievable SNR was limited by the low IR power at the sample and by 1/f noise associated with sub-kHz mechanical chopping [17].

Photothermal detection has also been combined with scanning probe microscopy. In atomic force microscopy–IR (AFM-IR) [18], absorbed IR radiation produces a localized thermal expansion of the sample that deflects an AFM cantilever, so that the deflection measures the absorbed energy. By overcoming the diffraction limit of IR light, AFM-IR combines the chemical specificity of vibrational spectroscopy with nanoscale spatial resolution, enabling chemical mapping of heterogeneous samples. AFM-IR has also been demonstrated in liquids for soft matter and biological samples [19,20], and the high sensitivity of scanning-probe photothermal techniques [2] has enabled measurements on individual crystals [21]. Zanni et al. developed Fourier-transform photothermal spectroscopy (FT-PTS), a time-domain implementation of AFM-IR that extends the spectral bandwidth to the visible region [22] and, by using ultrashort broadband pulses, opens a route to higher-order spectroscopies. Scanning-probe pump–probe spectroscopy has achieved sub-diffraction spatial resolution by recording tip–sample forces that respond to the material's nonlinear polarization [23,24]. In AFM-2DIR, a tip-enhanced femtosecond IR pulse sequence generates a localized photothermal response, and Fourier analysis of the cantilever response yields spatially resolved two-dimensional IR spectra, revealing vibrational coupling and energy-transfer pathways at the nanoscale [25]. Extensions of FT-PTS to two-dimensional electronic spectroscopy in the visible have also been proposed theoretically [26].

Here we present an alternative implementation of FT-PTS in the visible, using broadband supercontinuum excitation and TL detection in liquid samples. Interference between a pair of pump pulses allows absorption spectra to be retrieved without wavelength scanning, while TL detection provides high sensitivity and rapid acquisition, with each scan taking 16 s. We demonstrate the approach on phloxine B, a dye with a low fluorescence quantum yield, and retrieve its photothermal absorption spectrum at micromolar concentrations.

## 2. Method

### *2.1 Principle of Fourier-transform spectroscopy with photothermal detection*

Transient thermal-lens spectroscopy relies on detecting changes in the local refractive index of the solvent resulting from heat dissipated by the sample [13,27]. Upon resonant excitation by the pump laser, the sample is promoted to higher electronic states. For samples with low fluorescence quantum yields, a significant fraction of the absorbed energy is dissipated through non-radiative relaxation channels. The resulting heat produces a local temperature (T) increase, which in turn induces a change in the refractive index (n) through the thermo-optic coefficient ($\frac{dn}{dT}$) of the solvent. This localized refractive-index change acts as a transient lens in the sample. When a nonresonant photothermal probe beam propagates through the sample, it diverges upon passing through the transient lens formed by the photothermal response. This results in changes to the probe intensity transmitted through an iris placed after the sample, thereby encoding the photothermal response. For a non-fluorescent sample, the heat dissipated into the solvent is proportional to the absorbed optical energy, making the photothermal response an indirect measure of the sample absorptance from which the absorption spectrum can be found.

Typically, transient lens spectroscopy is performed with monochromatic pump laser. Here we extend transient lens detection to broadband excitation by combining linear Fourier Transform spectroscopy with transient lens detection. A common path TWINS interferometer [28] generates a pair of phase-stable supercontinuum pump pulses with a controllable relative temporal delay (τ). When the relative delay τ between the two pump pulses is scanned, the pump-pulse pair produces a delay-dependent spectral interference pattern. For two identical collinear pump pulses separated by a delay τ, the total pump energy at the sample can be expressed as

$$W_{\text{pump}}(\tau) \;\propto \int_{-\infty}^{\infty} |E(t) + E(t-\tau)|^2 \, dt \quad (1a)$$

$$= \int_{-\infty}^{\infty} |E(t)|^2 dt + \int_{-\infty}^{\infty} |E(t-\tau)|^2 dt \; + S(\tau) + S^*(\tau) \qquad (1b)$$

consisting of a constant background as well as the τ-dependent correlation function $S(\tau)$ where

$$S(\tau) \;= \int_{-\infty}^{\infty} E^*(t-\tau) E(t) \, dt \qquad (2)$$

By the Wiener-Khinchin theorem [29], $S(\tau)$ and the pump power spectrum $|E(\omega)|^2$ are a Fourier pair:

$$S(\tau) \;= \frac{1}{2\pi} \int_{-\infty}^{\infty} |E(\omega)|^2 \, e^{-i\omega\tau} d\omega \qquad (3)$$

This expression can be generalized to any time-invariant signal that depends linearly on the pulse energy

$$S(\tau) \;= \frac{1}{2\pi} \int_{-\infty}^{\infty} g(\omega) |E(\omega)|^2 \, e^{-i\omega\tau} d\omega. \qquad (4)$$

where $g(\omega)$ is a frequency domain function describing the response of the system to two electric field interactions, one from each pump pulse. The measured signal $W(\tau)$ is a time-integrated energy and therefore must be real for any given pump spectrum, which requires $g(\omega)$ to be a real quantity.

Here we wish to describe the delay-dependent photothermal signal produced when the pump-pulse pair is incident on the sample, leading to the absorption of pump energy by the sample. In this case, $g(\omega)$ becomes the sample absorptance, which we denote by $A_{\mathrm{PT}}(\omega)$:

$$\mathrm{S}_{PT}(\tau) \propto \int_{-\infty}^{\infty} A_{\mathrm{PT}}(\omega)|E(\omega)|^2\, e^{-i\omega\tau} d\omega \tag{5}$$

Thus Fourier transform of $\mathrm{S}_{PT}(\tau)$ gives the photothermal excitation spectrum $S_{\mathrm{PT}}(\omega)$ where

$$S_{\mathrm{PT}}(\omega) = C A_{\mathrm{PT}}(\omega)|E(\omega)|^2 \tag{6}$$

Here $C$ is a proportionality constant that includes the photothermal conversion efficiency, the probe detector response, lock-in gain, and Fourier-transform normalization. In general for a sample with fluorescence quantum yield of $\phi_f$, the photothermal conversion efficiency will be frequency dependent: $C(\omega) = C_0\left[1 - \phi_f \langle\omega_f\rangle/\omega\right]$, where $\langle\omega_f\rangle$ is the average fluorescence frequency, weighted by photon number [30]. This frequency dependence assumes that Kasha's rule holds and reflects the fact that higher frequency excitation populates higher lying states that yield more heat than lower frequency excitation via nonradiative relaxation within the excited state manifold. We note that we have neglected phosphorescence and photoproducts which could further reduce the photothermal conversion efficiency. For the case of negligible $\phi_f$, or when measuring relatively narrowband absorption bands such that $\langle\omega_f\rangle/\omega$ varies little across the band, we can neglect the frequency dependence of the photothermal conversion efficiency and make the approximation that $C(\omega) \approx C_0$. In this case the absorptance $A_{\mathrm{PT}}(\omega)$ can be determined by normalizing the photothermal excitation spectrum by the measured pump power spectrum $|E(\omega)|^2$.

$$A_{\mathrm{PT}}(\omega) \propto \frac{S_{\mathrm{PT}}(\omega)}{|E(\omega)|^2} \tag{7}$$

Finally, the photothermal absorptance spectrum $A_{\mathrm{PT}}(\omega)$ is related to the sample linear absorption spectrum $A(\omega)$, where $A(\omega) = -\,log_{10}\, I(\omega)/I_0(\omega)$ through the following relation:

$$A_{PT}(\omega) = 1 - 10^{-A(\omega)} \tag{8}$$

Thus for optically thin samples, the photothermal absorptance spectrum differs from the absorption spectrum only by a constant factor: $A_{PT}(\omega) \approx ln(10)A(\omega)$. In the case of higher absorbance, Eq. 8 must be used to correctly compare the lineshapes of $A(\omega)$ and $A_{PT}(\omega)$.

### *2.2 Experimental setup*

The schematic of the experimental setup is shown in Fig. 1(A). The 800 nm output of an 80 MHz mode-locked Ti-sapphire oscillator (Mai-Tai SP) passes through a Faraday isolator to prevent back-reflections and is divided into pump and probe arms. In the pump arm, a 20x microscope objective is used to focus 1.2 nJ pulses into a photonic crystal fiber (PCF) (Femtowhite800, NKT Photonics) to generate a supercontinuum that spans 500 – 1000 nm. A 750 nm short-pass filter limits the pump bandwidth to the desired range. A combination of a half-wave plate (AHWP05M-500) and linear polarizer (GTH5M-A) is used to control the input polarization and intensity of the pump laser coupled into the PCF. Dispersion introduced by Faraday isolator, polarization optics, and focusing objective is compensated using a prism-pair compressor placed before the PCF. The resulting pump supercontinuum is modulated at 10 kHz using a mechanical chopper (MCF10HP, MC2000B). A birefringent wedge-based TWINS interferometer [28] generates a pair of phase-stable pump pulses with a controllable temporal delay. A motorized actuator (LTA-HL) provides control of the pump delay (τ). The 800 nm fundamental output of the Ti: Sapphire oscillator is used as the non-resonant photothermal probe. A telescopic assembly of a 10 cm concave and 15 cm convex lens in the probe is used in controlling the axial separation between the pump and probe foci. For optimal photothermal signal intensity, the photothermal probe focuses before the pump focus [13]. A 20 cm concave mirror (CM508-200-P01, Thorlabs) focuses the pump and probe at the sample position with a crossing angle of 3.3 degrees with respective $1/e^2$ radii of 29.7 µm and 25.5 µm as measured by a CMOS camera (DCC1545M-GL).

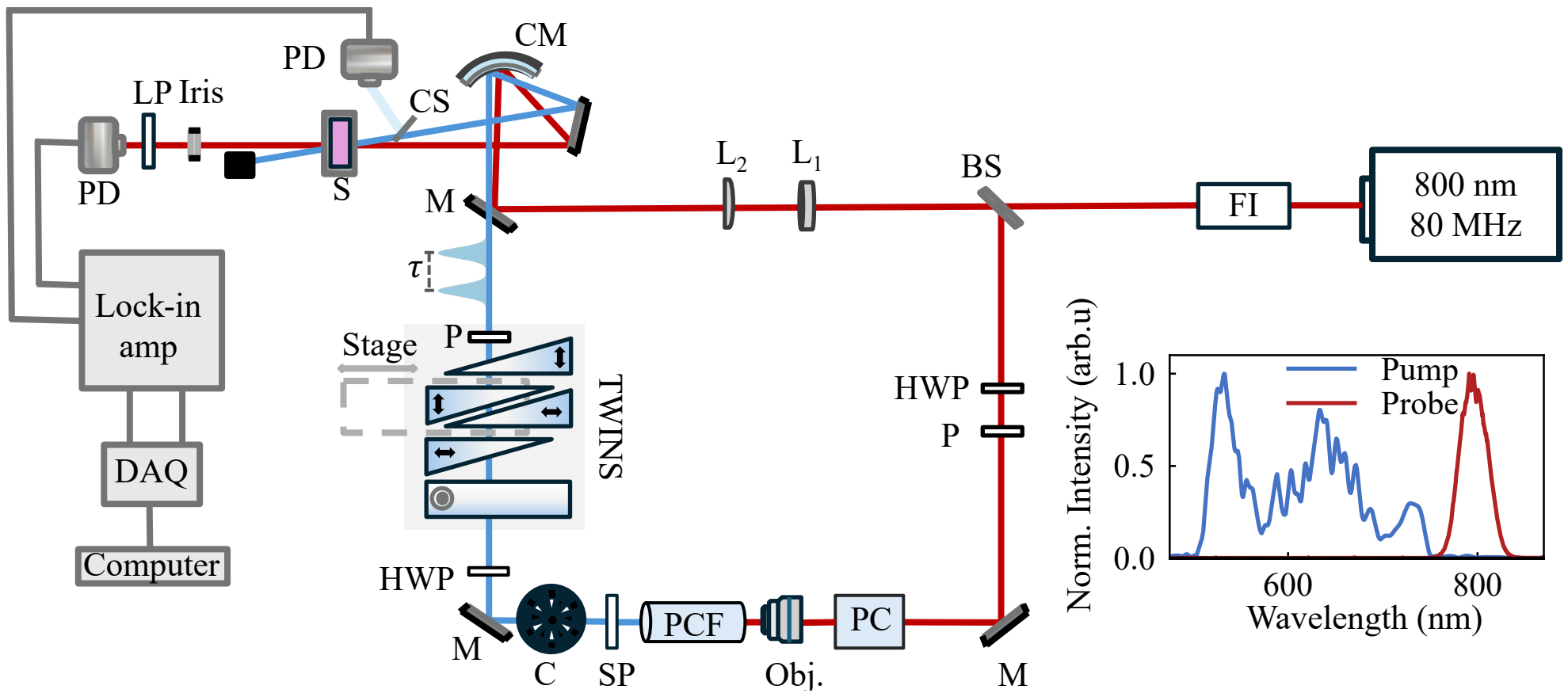

Fig. 1. (a) Experimental setup for Fourier transform photothermal spectroscopy with transient lens detection. FI: Faraday isolator; BS: Beam splitter; HWP: Half-wave plate; P: polarizer; M: mirror; PC: Prism compressor; Obj: objective; PCF: Photonic crystal fiber; SP: Shortpass filter; C: Chopper; L1: 10 cm concave lens; L2: 15 cm convex lens; CM: Curved mirror; CS: coverslip; S: Sample; LP: Longpass filter; PD: Photodiode. Normalized pump and probe spectra are shown in the inset.

After the sample the transmitted non-resonant probe is spatially filtered using an iris, which also acts as a beam block to the transmitted pump. The transmitted probe is detected by a silicon photodiode (PDA100A2, Thorlabs, Gain 0 dB, 11MHz BW), referred to as the signal photodiode. A 775 nm long pass filter is used to suppress any stray pump reflections reaching the signal photodiode. A small fraction of the pump beam is reflected from a coverslip placed before the sample and directed onto a second silicon photodiode referred to as the reference photodiode. The signal photodiode records the transmitted probe intensity, while the reference photodiode records the corresponding pump field autocorrelation from which the pump power spectrum can be determined. The transmitted probe and reference photodiode signals are demodulated at 10 kHz by a lock-in amplifier (HF2LI, Zurich Instruments) using a fourth-order low-pass filter with a 1 ms time constant. The resulting outputs from the lock-in are recorded simultaneously by a data acquisition card (DAQ, NI PCIe-6321) at 10 kHz.

### *2.3 Data acquisition*

Fig. 2 shows a schematic of the pulse sequence and the heating cycle. The pump beam is modulated at 10 kHz resulting in pump ON - OFF windows of 50 µs duration each. Each pump ON window consists of 4000 pump pulses. Hence, the sample molecules undergo excitation and relaxation multiple times during the heating window. The dissipation of the heat deposited following each excitation is however very slow. The characteristic thermal relaxation time ($\tau_c$ =

$\frac{r^2}{4\alpha}$) quantifies the timescale of heat diffusion across a length scale of r in a solvent of thermal diffusivity α. For a pump beam with a 1/e$^2$ radius of 29.7 µm, $\tau_c$ corresponds to 1.55 ms in water. Thus, heat accumulates during the ON window and dissipates slowly during the OFF window. After a few heating cycles, the heat accumulation and dissipation rates reach a steady state. This results in a constant heat offset with modulations corresponding to heating ON- OFF timescales as shown in Fig. 2. In the current implementation, a peristaltic pump is used to continuously flow the sample at a rate sufficient to displace the heated sample volume from the pump excitation region within a single ON–OFF modulation cycle, thereby improving the modulation contrast. The variation in the local refractive index in response to the heating cycle causes modulations in the probe intensity transmitted through the iris. Photothermal contrast is defined as $\frac{I_{OFF}-I_{ON}}{I_{OFF}}$, where $I_{ON}$ and $I_{OFF}$ are the transmitted probe intensities during the ON and OFF states of the heating cycle respectively. For maximum contrast the heat deposited during the ON window should be dissipated through thermal diffusion or sample flow during the OFF window. In the current experimental setting, contrast is ensured by the sample flow rate. Hence at faster flow rates, the experiment could potentially be performed at higher modulation frequencies for improved SNR and faster data acquisition. In contrast, for film samples or imaging applications, the modulation frequency would be limited by the characteristic thermal relaxation timescales. Photothermal IR imaging performed on PMMA films on silicon substrates, with thermal decay constant of 1.1 µs [31] suggests the feasibility of photothermal imaging of film samples at higher repetition rates.

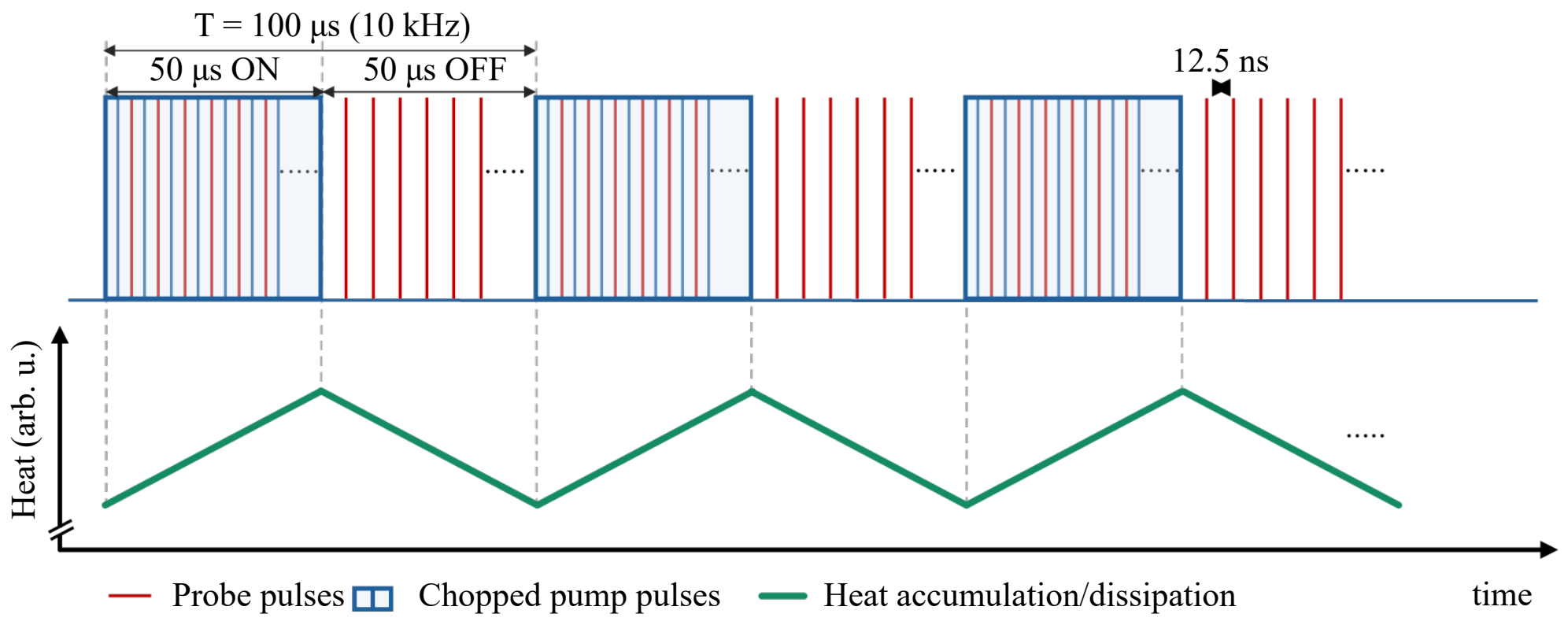

Fig. 2. (Top) A schematic of the pulse sequence. The red vertical lines represent the photothermal probe pulses at 80 MHz repetition rate. The pump beam is modulated at 10 kHz using a mechanical chopper. The transparent blue boxes represent the heating window with the blue vertical lines representing the pump pulses. One heating window has 4000 pump pulses. (Bottom) Schematic of heat accumulation and dissipation. In the 50 µs heating window, the heat deposited through the non radiative relaxation channels build up, followed by dissipation through thermal diffusion and sample flow in the next 50 µs. The break in the Y axis represents a constant heat offset.

### *2.4 Data processing*

For weakly absorbing samples we scan the delay forward and backward, recording multiple signal and reference interferograms to enable averaging to improve the signal-to-noise ratio of the measurement. Each interferogram is first offset corrected by subtracting its average intensity. The interferogram is then zero-padded to 2N+1 points and apodized using a tanh-edged window function to minimize edge artifacts, where N=80,000. Prior to averaging multiple scans, it is essential to ensure that the zero delay matches for each scan to avoid losing fringe contrast and therefore signal amplitude. Since both the signal and reference functions are even with respect to zero delay, the spectral phase will be exactly zero when $\tau = 0$ is correctly chosen and the resulting spectrum will be purely real. The zero delay can be determined via Fourier-transform of the reference interferogram, from which any residual $\Delta\tau$ can be found using a linear fit to the spectral phase $\phi(\omega) = \omega\Delta\tau$. Upon multiplication by $e^{-i\omega\Delta\tau}$ , inverse Fourier-transformation yields the time-domain zero-time-corrected signal. Use of the reference interferogram is valid because only the absorptive (real) part of the response deposits heat, so $S_{\mathrm{PT}}(\omega)$ carries no extra spectral phase. The SNR of the reference scan is typically higher than that of the signal, particularly at low concentration, making it a better choice for determining the zero delay of each scan. Prior to averaging, all backward-scan traces are reversed in time and all traces are averaged to obtain the final phase-corrected signal and reference interferograms. The final averaged time domain signal is Fourier transformed to obtain the final averaged photothermal excitation spectrum $S_{\mathrm{PT}}(\omega)$ and reference spectrum $S_{pump}(\omega)$ where the frequency axis is determined via calibration of the TWINS interferometer [28]. Eq. 7 shows that the photothermal absorptance spectrum $A_{\mathrm{PT}}(\omega)$ can be found by dividing the photothermal excitation spectrum $S_{\mathrm{PT}}(\omega)$ by the reference spectrum $S_{pump}(\omega)$. Note that here we make the approximation that $S_{pump}(\omega) \propto |E(\omega)|^2$ , which neglects the frequency dependence of both the coverslip reflectivity and the photodiode response.

## 3. Results and Discussion

We performed demonstration measurements on an aqueous solution of the xanthene dye phloxine B, which has an absorption maximum at 540 nm, with an absorbance of ~0.44 in a 200 µm path-length flow cell. Phloxine B is a halogenated derivative of fluorescein with four bromine atoms on the xanthene core and four chlorine atoms on the phenyl core. Halogenated derivatives show lower fluorescence quantum yields and higher non radiative decay rates relative to the fluorescein parent, with strongest quenching effect in aqueous solutions. In alkaline aqueous solution, fluorescein has a fluorescence quantum yield of 0.92, while for the brominated derivative eosin Y it drops to 0.2 in water [32]. Phloxine B has the same bromination pattern on the xanthene core as eosin Y and is therefore expected to be similarly weakly fluorescent in water with most of the absorbed energy released through non-radiative pathways.

The pump spectrum was selected to overlap with the phloxine B absorption band, although the PCF output can be extended further into the blue. The pump and probe pulse energies at sample position are 7.8 pJ and 250 pJ, corresponding to peak fluences of 0.56 µJ/cm$^2$ and 24.5 µJ/cm$^2$, respectively. The sample is continuously flowed through the flow cell using a peristaltic pump (Masterflex C/L 77122-24) at a flow rate of 28 ml/min. The signal and reference intensity are recorded simultaneously while the mechanical stage raster scans a pump delay of ~ +/- 200 fs. Each raster scan takes ~16 s and is limited by the maximum stage speed.

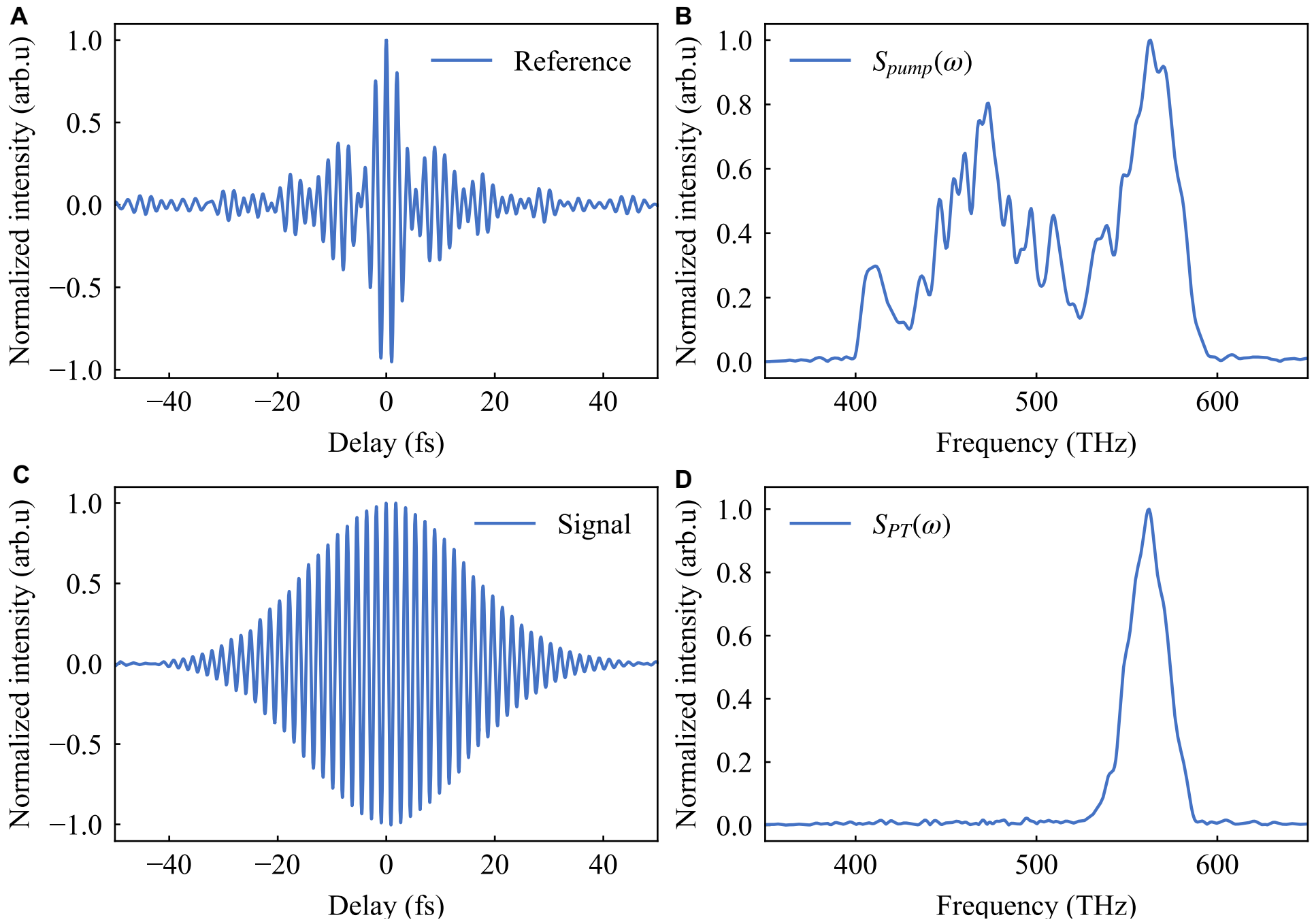

Fig. 3. (a) The pump intensity recorded in the reference photodiode as a function of pump delay τ and (b) its Fourier transform. (c) The transmitted photothermal probe intensity recorded as a function of pump delay τ. (d) Fourier transformed signal $S_{PT}(\omega)$.

Experimental results from Fourier transform photothermal absorption measurements are shown in Fig. 3. Fig. 3(A) shows the reference intensity as a function of the pump delay τ, while the corresponding Fourier transform spectrum $S_{pump}(\omega)$ is shown in Fig. 3(B). Fig. 3(C- D) correspond to photothermal signal detection. The photothermal signal is plotted as a function of the pump delay in Fig. 3(C). For a weakly fluorescent sample, the modulations in the transmitted probe intensity encode the sample absorptance information. Fourier Transform of the signal gives the photothermal excitation spectrum $S_{PT}(\omega)$ shown in Fig. 3(D). The SNR is determined from the normalized spectrum and the standard deviation of the noise floor within the 450 – 500 THz range. Owing to the high sensitivity of transient lens spectroscopy a single scan (forward and backward) of 16 s yields an SNR of 210.

Figure 4 shows the photothermal absorptance spectrum determined from the measured photothermal excitation spectrum and reference spectrum of Fig.3. For comparison we also show the reference absorptance spectrum of phloxine B, measured by a commercial UV-visible spectrometer (CCS200, Thorlabs) and corrected for the relatively high absorbance of 0.44. The dip in the photothermal absorptance spectrum in the blue side of the spectrum ~510 nm is caused by the falling edge of the pump spectrum. Similarly, the weak shoulder at 570 nm corresponds to the dip in the intensity of the pump spectrum at the same wavelength, increasing the error in the measurement at that location. Overall, the absorptance spectrum reconstructed from the photothermal signal shows good agreement with the phloxine B absorptance measured in a spectrophotometer. We note that our derivation of Eq. 7 assumed a constant photothermal conversion efficiency. With $\phi_f \approx 0.2$ and $\langle \lambda_f \rangle \approx 565$ nm [33], $1 - \phi_f \cdot \lambda / \langle \lambda_f \rangle$ changes by ~2–3% between 500 and 570 nm, making our assumption of a constant photothermal conversion efficiency reasonable.

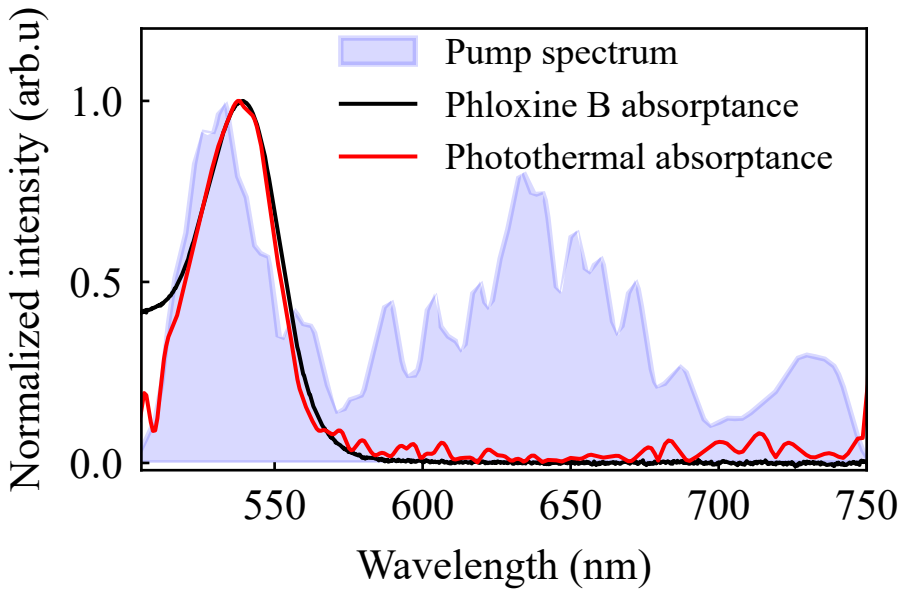

Fig. 4. Photothermal absorptance spectrum of phloxine B (black) measured from a single scan overlaid with the absorptance spectrum determined from a commercial UV-visible spectrometer (red). The pump spectrum is shown in the background.

To test the sensitivity of transient lens detection, we performed measurements on phloxine B solutions at different concentrations. The sample absorbance was sequentially decreased from 0.44 (470 μM) to 0.22 (235 μM), 0.025 (26.7 μM) and 0.013 (13.9 μM) corresponding to ~ 34-fold reduction in concentration. The absorption spectra were measured in a 200 μm cuvette using a commercial UV-visible spectrometer (CCS200, Thorlabs). The extinction coefficient for phloxine B in water was measured to be $(4.7 \pm 0.1) \times 10^4$ $M^{-1}cm^{-1}$. Fig. 5(A) compares the normalized photothermal absorptance spectra obtained from a single scan for the four sample concentrations. Note that the 470 μM and 235 μM samples show consistent peak positions with the lower concentration spectra but are broader. This is a consequence of the nonlinearity of Eq. 8 at higher absorbance values. The noisy spectral features observed between 560 – 650 nm arise from dividing the relatively flat signal intensity in this spectral window with the structured laser spectrum (Fig. 3(B,D)). At lower concentrations, the signal decreases proportionally, resulting in a reduced SNR. As a result, noise features distort the photothermal absorptance peak in measurements on dilute samples. The spectral artefacts between 560 and 650 nm also become more prominent, as observed in Fig. 5(A).

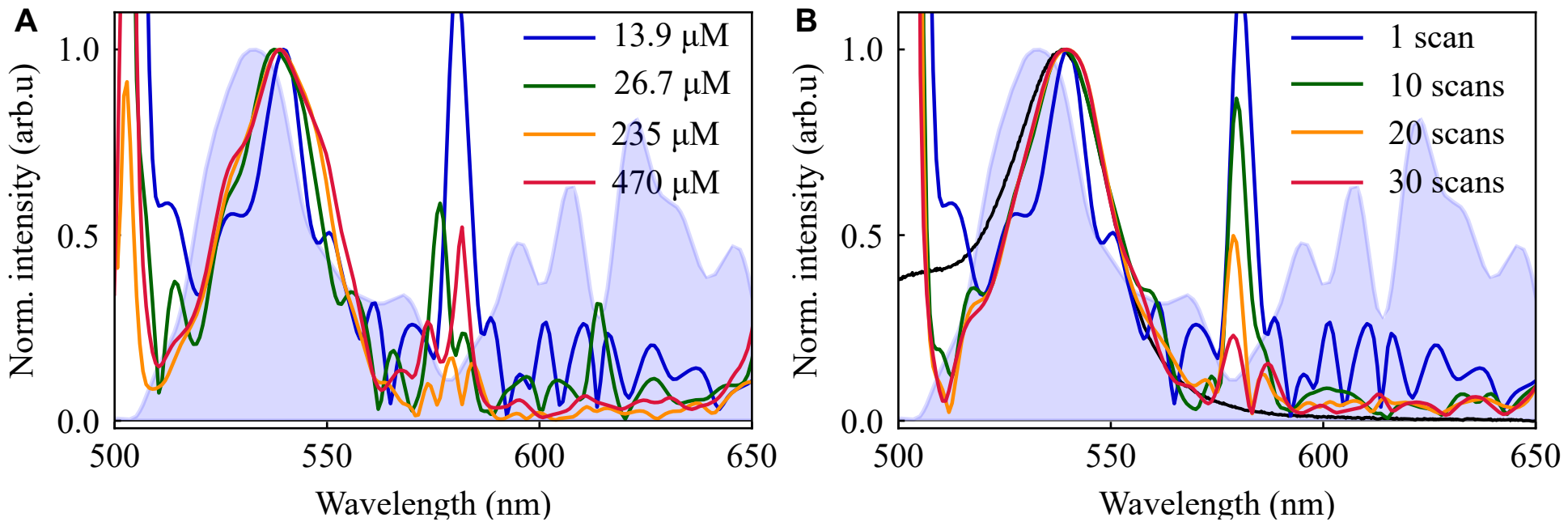

Fig. 5. (a) Normalized photothermal absorptance spectra of phloxine B from a single scan at different concentrations. (b) Normalized photothermal absorptance spectra of 13.9 μM phloxine B for different number of averages. The SNR improves with increasing number of scans owing to noise-floor suppression through averaging. The pump spectrum is shown in the background in both panels. The black trace shows the phloxine B absorption spectrum measured with a commercial UV-visible spectrometer.

For the 13.9 μM sample, Fig. 5(B) shows improvement in the retrieved photothermal absorptance spectrum with increasing the number of scans. Averaging over 10 scans significantly reduces the spectral distortions resulting in a spectrum that shows good agreement with the reference absorption spectrum (black). The acquisition time for 10 scans is ~2.6 minutes. Further increasing the number of scans provides additional noise floor suppression and improves the SNR as shown in Fig. 5(B). The SNR could be enhanced further by combining faster delay-stage scanning with multiple raster scans, and by using a higher stability light source. These results demonstrate that transient lens detection enables the retrieval of photothermal absorptance spectra at low sample concentrations, with the SNR further improved through signal averaging. Previous studies on noise analysis in rapid scan nonlinear spectroscopy experiments have demonstrated improved suppression of 1/f noise through higher modulation frequencies, faster delay scans and multiple raster scans, compared to a single slowly acquired and heavily averaged scan. These findings suggest that the SNR could be further enhanced in transient lens spectroscopy by employing higher modulation frequencies, faster delay stages and shorter acquisition times [34,35].

## Conclusion

We have demonstrated Fourier-transform photothermal absorptance spectroscopy in the visible spectral region using a broadband excitation source. Photothermal measurements on phloxine B establish the feasibility of this approach for studying samples with low fluorescence quantum yields. The sensitivity of the technique was evaluated using micromolar concentrations, demonstrating its capability to retrieve photothermal absorptance spectra from dilute samples. The high sensitivity of transient lens detection enables rapid measurements with short acquisition times, while signal averaging provides further SNR improvement for weak signals. The experimental design is not subject to intrinsic bandwidth limitations, opening up possibilities for extending this approach to widefield imaging and higher-order spectroscopic measurements.

### Funding.

U.S. Department of Energy (DE-SC0010575, DE-SC0016384)

Natural Sciences and Engineering Research Council of Canada (RGPIN-2025-06687).

### Acknowledgment.

We gratefully acknowledge funding from the U.S. Department of Energy (No. DE-SC0010575 to A.S.T and DE-SC0016384 to M.Z.). J. P. O., T. H. and L. Y. acknowledge support from the Natural Sciences and Engineering Research Council of Canada (NSERC) (grant RGPIN-2025-06687).

### Disclosures.

The authors declare no conflicts of interest.

### Data availability statement.

The data that support the findings of this study are openly available in Zenodo [https://doi.org/10.5281/zenodo.24289526].